\def\year{2021}\relax

\documentclass[letterpaper]{article} 
\usepackage{aaai21}  
\usepackage{times}  
\usepackage{helvet} 
\usepackage{courier}  
\usepackage[hyphens]{url}  
\usepackage{graphicx} 
\usepackage{graphicx}  
\usepackage{natbib}  
\usepackage{caption} 

\usepackage{amsmath}
\usepackage{amssymb}
\usepackage{amsfonts}
\usepackage{accents}
\usepackage{trimclip}
\usepackage{mathdots}
\usepackage{booktabs}
\usepackage{bm}
\usepackage{multirow}
\usepackage{subfigure}
\usepackage[export]{adjustbox}

\usepackage[capitalize]{cleveref}
\Crefname{equation}{Eq.}{Eqs.}

\usepackage{algorithmic}
\usepackage{textcomp}
\usepackage{xcolor}
\usepackage[caption=true,font=footnotesize]{subfig}
\usepackage{enumitem}
\usepackage{threeparttable}
\usepackage{comment}

\DeclareGraphicsExtensions{.pdf,.png,.jpg}

\newtheorem{defn}{Definition}
\newtheorem{prbl}{Problem}

\newcommand{\ourmodel}{{\sc TrainOR}}
\newcommand{\bjsh}{{\bf BJ\ensuremath{\rightarrow}SH}}
\newcommand{\shhz}{{\bf SH\ensuremath{\rightarrow}HZ}}
\newcommand{\gzfs}{{\bf GZ\ensuremath{\rightarrow}FS}}

\makeatletter
\newcommand{\printfnsymbol}[1]{%
  \textsuperscript{\@fnsymbol{#1}}%
}
\makeatother

\title{
Out-of-Town Recommendation with Travel Intention Modeling
}
\author{
	Haoran Xin,\textsuperscript{\rm 1, 2}
	Xinjiang Lu,\textsuperscript{\rm 2, 3}\thanks{Corresponding author. This work was done when Haoran Xin was an intern at the Baidu Research.}
	Tong Xu,\textsuperscript{\rm 1}
	Hao Liu,\textsuperscript{\rm 2, 3}
	Jingjing Gu,\textsuperscript{\rm 4}
	Dejing Dou,\textsuperscript{\rm 2, 3}
	Hui Xiong\textsuperscript{\rm 5}
    \\
}
\affiliations{

\textsuperscript{\rm 1}University of Science and Technology of China, 
\textsuperscript{\rm 2}Business Intelligence Lab, Baidu Research\\
\textsuperscript{\rm 3}National Engineering Laboratory of Deep Learning Technology and Application, China\\
\textsuperscript{\rm 4}Nanjing University of Aeronautics and Astronautics, 
\textsuperscript{\rm 5}Rutgers University\\
xinhaoran@mail.ustc.edu.cn,  \{luxinjiang, liuhao30, doudejing\}@baidu.com, tongxu@ustc.edu.cn, gujingjing@nuaa.edu.cn, hxiong@rutgers.edu
}

\begin{document}
\maketitle

\begin{abstract}
Out-of-town recommendation is designed for those users who leave their home-town areas and visit the areas they have never been to before. 
It is challenging to recommend Point-of-Interests (POIs) for out-of-town users since the out-of-town check-in behavior is determined by not only the user’s home-town preference but also the user’s travel intention. 
Besides, the user’s travel intentions are complex and dynamic, which leads to big difficulties in understanding such intentions precisely. 
In this paper, we propose a {\sc Tra}vel-{\sc in}tention-aware {\sc O}ut-of-town {\sc R}ecommendation framework, named {\sc TrainOR}.
The proposed {\sc TrainOR}~framework distinguishes itself from existing out-of-town recommenders in three aspects. 
First, graph neural networks are explored to represent users’ home-town check-in preference and geographical constraints in out-of-town check-in behaviors. 
Second, a user-specific travel intention is formulated as an aggregation combining home-town preference and generic travel intention together, where the generic travel intention is regarded as a mixture of inherent intentions that can be learned by Neural Topic Model (NTM). 
Third, a non-linear mapping function, as well as a matrix factorization method, are employed to transfer users’ home-town preference and estimate out-of-town POI’s representation, respectively. 
Extensive experiments on real-world data sets validate the effectiveness of the {\sc TrainOR}~framework.
Moreover, the learned travel intention can deliver meaningful explanations for understanding a user’s travel purposes.
\end{abstract}

\begin{figure*}[t]
  \centering
  \includegraphics[width=.9\textwidth]{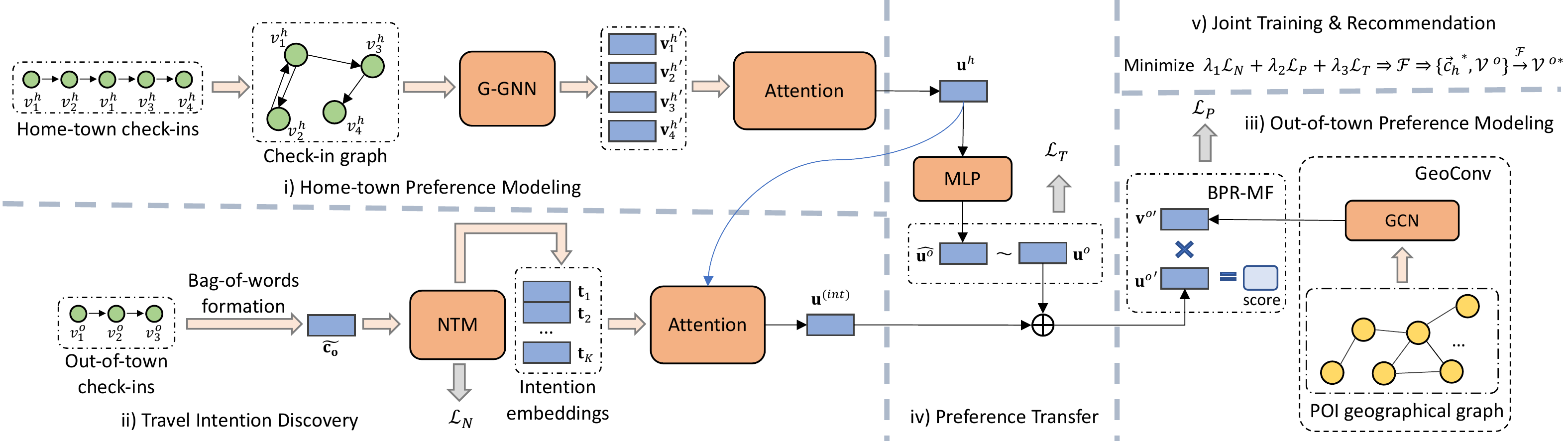}
  \caption{The overview of \ourmodel~framework.}
  \label{fig:framework}
\end{figure*}

\section{Introduction}

Point-of-Interest (POI) recommendation is an important task in location-based services (LBS),  which tends to act as a more pivotal part in people's daily life. 
Recently, since the POI check-in data having accumulated rapidly over time, a more refined recommendation problem, out-of-town recommendation, is coming into focus. 
To be specific, out-of-town recommendations are designed for those users who travel from their home-town areas to out-of-town areas they have seldom been to before. 

Out-of-town recommendation problem suffers from the cold-start issue a lot due to the insufficiency of out-of-town check-ins~\cite{ference2013location}.
Traditional POI recommender systems (POI RSs) fail to make appropriate recommendations to tackle such severe cold start issues.
The reasons are: 
1) Individual's home-town preferences cannot be used for out-of-town recommendations directly due to the gap between home-town preferences and out-of-town behaviors (i.e. interest drifts); 
and 2) The travel intention, which tends to affect the out-of-town check-in behaviors, is often ignored in these POI RSs. 

In the literature, some research efforts have been made to attack the out-of-town recommendation problem.
For instance, 
\cite{pham2017general} recommends out-of-town region of POIs instead of individual POIs by exploiting the proximity of human mobility. 
\cite{ference2013location} proposes a recommender for out-of-town users by taking into account user preference, social influence and geographical proximity.
Besides, some researchers have also paid attentions to interest drifts when addressing the out-of-town recommendation problem~\cite{yin2014lcars,yin2016joint,wang2017location}.
However, none of these approaches comprehensively integrate users' preferences, interest drifts and complex travel intentions as a whole. 

To this end, in this paper, we propose a {\sc Tra}vel-{\sc in}tention-aware {\sc O}ut-of-town {\sc R}ecommendation framework, named \ourmodel. 
Specifically, 
we first devise a user's preference representation module based on Gated Graph Neural Network (G-GNN) to explore the underlying structural information encoded in user's home-town check-ins. 
After being aggregated via an attention network, the user's home-town preference is further transferred into out-of-town preference through a non-linear mapping function, i.e. multi-layer perceptron (MLP).
In this way, the interest drifts from home-town to out-of-town can be captured directly. 
Besides, we devise a travel intention discovery module by developing a Neural Topic Model (NTM) followed by user-specific travel intention aggregation. 
In particular, we assume that each out-of-town check-in activity can be drawn from a latent topic mixture which can be further generated by Gaussian Softmax construction, then we adopt variational inference to uncover users' generic travel intention without extra supervision. 
Moreover, the aforementioned user's home-town preference is integrated into the disclosed generic travel intention to generate user-specific travel intention via another attention network. 
In addition, we represent user's out-of-town preference by exploiting a matrix factorization~(MF) approach and enrich such out-of-town preference by taking into account the geographical proximity among out-of-town POIs.
Finally, a joint learning method is employed in an end-to-end manner to yield the trained recommender.
To sum up, our major contributions are as follows:
\begin{itemize}
    \item 
    We study the out-of-town recommendation problem by modeling user's complex travel intention. 
    \item
    We devise a framework \ourmodel~which is able to capture the user's home-town preference, user's interest drift from home-town to out-of-town, out-of-town geographical influence and user's travel intention comprehensively. 
    \item
    We demonstrate the effectiveness of \ourmodel~quantitatively and qualitatively through extensive experiments.
\end{itemize}

\section{Problem Definition} \label{sec:pre}

In this section, we formally define the out-of-town recommendation problem. 
We start by defining several concepts.

\begin{defn}[POI]
A POI is a spatial item related to a geographical location.
We use $v$ to represent a POI identifier.
\end{defn}

\begin{defn}[Check-in]
A user's check-in activity $c$ is represented by a three-tuple $ (u, t, v) $ which indicates that a user $u$ visits POI $v$ at timestamp $t$.
\end{defn}

\begin{defn}[User Home-Town]
Given a user $u$, we denote a region $ \tilde{r}_u $ as the user's home-town where the user lives in for a period of time, say, 6 months.
\end{defn}

\begin{defn}[Travel Behavior]
Given a user $u$, his/er travel behavior is represented by a five-tuple $ \tau = (u, \vec{c}_{h}, \vec{c}_{ o}, \tilde{r}_u, r_o ) $ which indicates that the user $u$ travels from his/er home-town $ \tilde{r}_u $ to out-of-town $ r_o $ and leaves check-in records in both home-town and out-of-town, which are represented by $ \vec{c}_{ h} $ and $ \vec{c}_{ o} $, respectively. 
\end{defn}

When a user $u$ travels from his/er home-town  $ \tilde{r}_u $ to an out-of-town $ r_o $, we take $u$ as an out-of-town user and aim to recommend a list of POIs located at $ r_o $ that $u$ may be interested in.
Formally, we have the following problem statement: 

\begin{prbl}[Out-of-town Recommendation]
given a set of users $\mathcal{U}$ who live in $\tilde{r}$, 
a target region $ r_o $, 
a set of out-of-town POIs $\mathcal{V}^o$ in $r_o$, 
and the travel behavior records $ \mathfrak{T} $ generated by $ \mathcal{U} $ when traveling from $\tilde{r}$ to $r_o$,
learn a function $ \mathcal{F}( \cdot ) $ by exploring $ \mathfrak{T} $ and $ \mathcal{V}^o $.
Then, recommend a list of POIs $\mathcal{V}^{o{\ast}} \subset \mathcal{V}^o $ to a new coming user $ u^{\ast} \notin \mathcal{U} $ given his/er home-town check-ins $ {\vec{c}_h}^{\ast} $ observed in  $\tilde{r}$: 
$ \{ {\vec{c}_h}^{\ast}, \mathcal{V}^o \} \xrightarrow{ \mathcal{F} } \mathcal{V}^{o{\ast}}  $.
\end{prbl}

\section{The Proposed Approach}

\subsection{Framework Overview}
We first present the overview of \ourmodel~framework which is illustrated in~\cref{fig:framework}.
The \ourmodel ~framework consists of five components: 

\begin{itemize}
    \item 
    {\bf Home-town preference modeling} 
    takes user's home-town check-ins as input and assigns a $d$-dimensional embedding to each of the visited POIs. 
    Then the user's home-town preference is encoded and aggregated by adopting G-GNN model and attention network.
    \item 
    {\bf Travel intention discovery} 
    takes the user's visited POIs in out-of-town as input in bag-of-words, and then an NTM model takes such input to discover the generic travel intention.
    Afterward, another attention network is adopted to summarize user-specific intention by integrating discovered intention and user's home-town preference. 
    \item 
    {\bf Out-of-town preference modeling} 
    assigns another two $d$-dimensional embeddings to each user and  out-of-town POI, and utilizes MF to learn the latent representations of users and POIs.
    Moreover, to model the geographical influence of POIs, a GeoConv is explored to process the geo-information bundled with POIs. 
    \item 
    {\bf Preference transfer} 
    receives the home-town preference embedding and captures the non-linear relationship from home-town to out-of-town via an MLP. 
    \item 
    {\bf Model learner} jointly minimizes the intention inference loss, preference estimation loss, and preference transfer loss to output the trained recommender $ \mathcal{F} $.
\end{itemize}

\subsection{Home-town Preference Modeling}
To encode users' home-town preference, we represent the structural information with the G-GNN model~\cite{wu2019session,li2015gated}.

\begin{figure}[t]
  \centering
  \includegraphics[width=0.875\columnwidth]{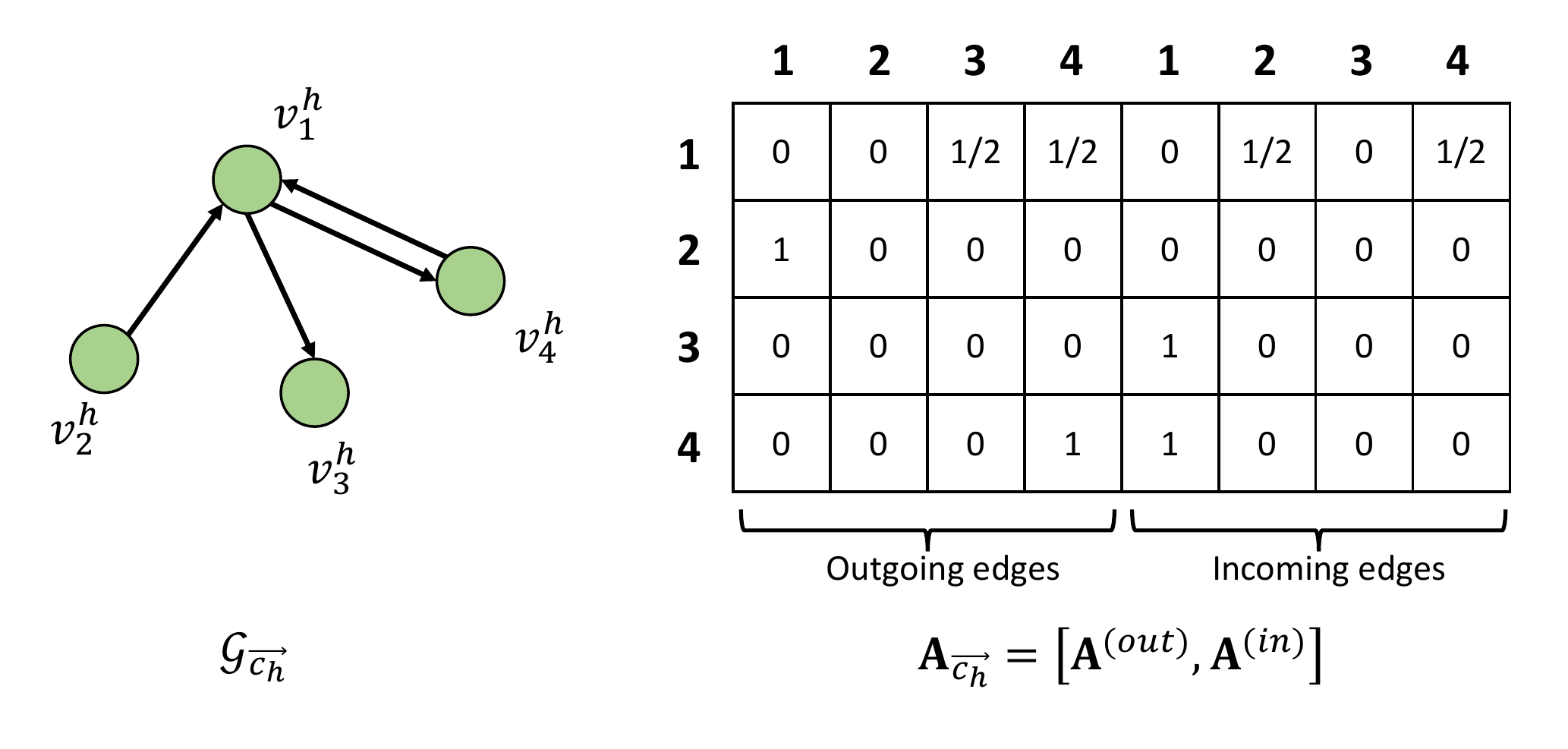}
  \caption{An illustration of check-in graph $\mathcal{G}_{\vec{c}_h}$ and the construction of corresponding $\mathbf{A}_{\vec{c}_h} = [ \mathbf{A}^{(out)}, \mathbf{A}^{(in)} ] $.}
  \label{fig:adjacent_matrix}
\end{figure}

Given a user $u$ and his/er home-town check-ins $\vec{c}_h$, we first build a directed graph $\mathcal{G}_{\vec{c}_h} = \left(\mathcal{V}_{\vec{c}}, \mathcal{E}_{\vec{c}}\right)$, 
where $\mathcal{V}_{\vec{c}} $ denotes the set of home-town check-ins 
and each pair of adjacent check-ins is represented by $(v^h_{i-1}, v^h_i) \in \mathcal{E}_{\vec{c}}$ ($v_i^h \in \vec{c}_h$). 
Notably, duplicated pairs of spatial items may exist in $\vec{c}_h$, we normalize all the weights of the edges in $\mathcal{G}_{\vec{c}_h}$.
Then, we construct the adjacent matrix $\mathbf{A}_{\vec{c}_h}$ (refer to~\cref{fig:adjacent_matrix}). 
The matrix $\mathbf{A}_{\vec{c}_h} \in \mathbb{R}^{D_1 \times 2D_1}$~(NOTE: $D_1 = |\mathcal{V}_{\vec{c}}|$) determines how spatial items communicate with each other via user's check-ins.
Next, we assign a $d$-dimensional embedding $\mathbf{v}_i^h$ to each vertex $v^h_i$ in $\mathcal{G}_{\vec{c}_h}$ and feed the corresponding embeddings $\mathbf{V}^h = \left(\mathbf{v}^h_1, \mathbf{v}^h_2, \ldots, \mathbf{v}^h_{D_1}\right)$ into the G-GNN.
$\forall v \in \mathcal{V}_{\vec{c}}$, the network propagates as follows:
\begin{equation} \label{e:ggnn:a}
  \mathbf{a}_{v}^{(t)} = \mathbf{A}_{ v: }^{\mathrm{T}} 
  						\left[ {\mathbf{v}_1^h}^{(t-1)}, {\mathbf{v}_2^h}^{(t-1)}, \cdots, 
  						  {\mathbf{v}_{ D_1 }^h}^{(t-1)}
  					   \right]^{\mathrm{T}} 
  					  + \bm{b}^g  ,
\end{equation}
\begin{equation} \label{e:ggnn:z}
  \mathbf{z}_v^{(t)} = \zeta \left( \mathbf{W}^z \mathbf{a}_v^{(t)} + \mathbf{U}^z {\mathbf{v}_v^h}^{(t-1)} \right)  ,
\end{equation}
\begin{equation} \label{e:ggnn:r}
  \mathbf{r}_v^{(t)} = \zeta \left( \mathbf{W}^r \mathbf{a}_v^{(t)} + \mathbf{U}^r {\mathbf{v}_v^h}^{(t-1)} \right)  ,
\end{equation}
\begin{equation} \label{e:ggnn:vtilde}
  \widetilde{{\mathbf{v}_v^h}^{(t)}}
  							= \tanh{ \left[ \overline{\mathbf{W}} \, \mathbf{a}_v^{(t)} + \overline{\mathbf{U}} 
  												\left(		\mathbf{r}_v^{(t)} \odot {\mathbf{v}_v^h}^{(t-1)} 	\right)	
  										\right]} ,
\end{equation}
\begin{equation} \label{e:ggnn:v}
  {\mathbf{v}_v^h}^{(t)} = (1 - \mathbf{z}_v^{(t)}) \odot {\mathbf{v}_v^h}^{(t-1)} 
  										 + \mathbf{z}_v^{(t)}   \odot
  						  \widetilde{{\mathbf{v}_v^h}^{(t)}} ,
\end{equation}
where $\mathbf{A}_{v:}$ are the two columns of blocks in $\mathbf{A}^{(out)}$ and $\mathbf{A}^{(in)}$ corresponding to $v$, and $\zeta(\cdot)$ is the sigmoid function.
In particular, \cref{e:ggnn:a} is the step that passes information between different POIs based on $\mathcal{G}_{\vec{c}_h}$. 
\cref{e:ggnn:z,e:ggnn:r,e:ggnn:vtilde,e:ggnn:v} are the update steps similar to GRU~\cite{cho2014properties}.
The updated embeddings learned by G-GNN are denoted as 
${\mathbf{V}^h}^{\prime} = \left[ {\mathbf{v}^h_1}^{\prime}, {\mathbf{v}^h_2}^{\prime}, \cdots, {\mathbf{v}^h_{D_1}}^{\prime} \right] $.

Furthermore, to summarize user's home-town preference, we adopt an attention network as follows:
\begin{equation} \label{e:home_preference}
    \begin{split}
        &\alpha_i = \mathbf{q}^{\mathrm{T}}
        \zeta\left(\mathbf{W}^p {\mathbf{v}^h_i}^{\prime} + \bm{b}^p \right)  ,
        \\
        &\mathbf{u}^h = \sum_{i=1}^{D_1}
        \alpha_i {\mathbf{v}^h_i}^{\prime}  ,
    \end{split}
\end{equation}
where $\mathbf{q} \in \mathbb{R}^d$ and $\mathbf{W}^p \in \mathbb{R}^{d \times d}$ weigh the home-town POIs, 
and $\mathbf{u}^h$ is the user's home-town preference embedding.

\subsection{Travel Intention Discovery}
Understanding travel intentions plays an important role in out-of-town recommendation.
Inspired by \cite{miao2017discovering,srivastava2017autoencoding}, we develop a Neural Topic Model (NTM) to uncover the inherent travel intentions without extra supervision. 

\subsubsection{Uncovering Generic Travel Intentions.}

Assume that each out-of-town check-in is generated by a latent topic mixture $\Theta \in \mathbb{R}^K$, which can be regarded as the generic travel intention mixture of users, where $K$ denotes the number of generic intentions.
Then, $\forall i$ ($1 \leq i \leq K$), we adopt an embedding $\mathbf{t}_i \in \mathbb{R}^d$ to represent the $i$-th travel intention. 
Afterward, given the out-of-town POI embedding matrix $\mathbf{E} \in \mathbb{R}^{|\mathcal{V}^o| \times d}$, 
the i-th generic out-of-town travel intention distribution over the out-of-town POIs, denoted as $\Phi_i$, can be determined as follows:
\begin{equation}\label{e:phi}
    \Phi_i = \mathrm{softmax}\left(\mathbf{E}\mathbf{t}_i\right)  ,
\end{equation}
where $\Phi_i \in \mathbb{R}^{ |\mathcal{V}^o| }$. 
Then we denote the whole out-of-town intention-POI distribution as $\Phi=\left(\Phi_1,\Phi_2,\ldots,\Phi_{K}\right)^\mathrm{T}$.

We assume that the distribution $\Theta$ can be generated by Gaussian Softmax construction. 
Let $\widetilde{\mathbf{c}_o} \in \mathbb{R}^{|\mathcal{V}^o|}$ be the bag-of-words vector to represent the user's out-of-town check-ins, then the generation of $\widetilde{\mathbf{c}_o}$ can be conducted as follows:
\begin{itemize}
    \item 
    Draw a latent variable $\mathbf{z}$ from a standard Gaussian distribution: 
    $\mathbf{z} \sim \mathcal{N}\left(\mathbf{0},\mathbf{I}\right)$.
    
    \item
    Generate the out-of-town intention distribution $ \Theta $ : 
    $\Theta = \mathrm{softmax} \left(F_{\Theta}\left(\mathbf{z}\right)\right)$, where $F_{\Theta}$ is a fully connected layer.
    
    \item
    For the i-th POI in $\widetilde{\mathbf{c}_o}$, draw a POI $v_i \sim {\Phi}^\mathrm{T} \, \Theta $.
\end{itemize}

As shown above, we can find that $p(\mathbf{z}) = \mathcal{N}\left(\mathbf{0}, \mathbf{I}\right)$. In order to make $\mathbf{z}$ traceable, a variational posterior distribution is introduced as below:
\begin{equation}\label{e:variational}
    q(\mathbf{z}|\widetilde{\mathbf{c}_o}) =
    \mathcal{N}\left(\bm{\mu},\bm{\sigma}^2 \right) ,
\end{equation}
where $\bm{\mu}$, $\bm{\sigma}^2$ are two prior parameters determined by the input bag-of-words vectors:
\begin{equation}\label{e:mu_sigma}
    \begin{split}
        \bm{\mu} = 
        F_{\mu}\left(F_{enc}\left(\widetilde{\mathbf{c}_o}\right)\right)  ,
        \\
        \bm{\sigma}^2 =
        F_{\sigma}\left(F_{enc}\left(\widetilde{\mathbf{c}_o}\right)\right) ,
    \end{split}
\end{equation}
where $F_{\mu}$, $F_{\sigma}$ are two multi-layer perceptrons (MLP) and $F_{enc}$ is an encoder layer which accepts bag-of-words inputs extracted from out-of-town check-ins.

As the neural variational inference {instructs}, we would like to maximize the variational lower bound. 
Thus, the intention inference loss is defined as follows:
\begin{equation}\label{e:ntm_loss}
    \begin{split}
        \mathcal{L}_N = & - \sum _ {u \in \mathcal{U}} \Big[\mathbb{E}_{q(\mathbf{z}|\widetilde{\mathbf{c}_o})}
        \left(\widetilde{\mathbf{c}_o}^\mathrm{T}
        \log{\left(\Phi^\mathrm{T}\Theta\right)}\right)
        \\
        & + \mathbb{D}_{\mathrm{KL}}
        \left(q\left(\mathbf{z}|\widetilde{\mathbf{c}_o}\right)
        ||p\left(\mathbf{z}\right)\right)\Big] ,
    \end{split}
\end{equation}
where $\mathbb{D}_{\mathrm{KL}}$ is the Kullback-Leibler divergence.

By optimizing the above loss, the generic travel intentions can be discovered without extra supervision.

\subsubsection{Summarizing User-Specific Travel Intention.}
Previous works~\cite{zeng2018topic,wei2019modeling} have paid attention to integrating the topic knowledge with downstream tasks. 
Inspired by these, we further design an attention network to probe the  dynamic travel intentions of users, which can explore intention knowledge according to user's home-town preference.

Specifically, after the generic out-of-town intention 
$\mathbf{T} = 
\left(
{\mathbf{t}_1,\mathbf{t}_2,\ldots,\mathbf{t}_{K}}
\right)^{\mathrm{T}}$ being acquired with NTM, we implement the attention network as follows:
\begin{equation}\label{e:intention_embedding}
    \begin{split}
        & \beta_i = \mathrm{softmax}\left(
        \mathbf{t}_i^\mathrm{T} 
        \mathbf{W}^t
        \mathbf{u}^h
        \right)  ,
        \\
        & \mathbf{u}^{(int)} = \sum_{i=1}^{K}
        \beta_i \mathbf{t}_i  ,
    \end{split}
\end{equation}
where $\mathbf{W}^t \in \mathbb{R}^{d \times d}$ is a trainable transition matrix. 
By fitting the user's preference, the {user-specific} intention embedding $\mathbf{u}^{(int)}$ can be aggregated adaptively.

\subsection{Out-of-town Preference Modeling} 
Geographical influence underlying out-of-town POIs is helpful in understanding users' out-of-town check-in behaviors. 
On the other hand, with the logged travel records $ \mathfrak{T} $, we can further enrich the representations of out-of-town POIs via exploiting the interactions between POIs and users.

Specifically, 
we first assign another $d$-dimensional embedding to each out-of-town POI denoted as $\mathbf{v}^o \in \mathbb{R}^d$,
and we have $\mathbf{V}^o = {\left(\mathbf{v}^o_1, \mathbf{v}^o_2, \ldots, \mathbf{v}^o_{D_2}\right)}^{\mathrm{T}}$ 
where 
$\mathbf{V}^o \in \mathbb{R}^{D_2 \times d}$~(NOTE: $D_2 = |\mathcal{V}^o|$).

Then, we build an undirected graph $\mathcal{G}_{geo} = (\mathcal{V}^o, \mathcal{E}^o)$ based on the geographical relations among POIs, 
and the edge $ e^o_{i, j} \in \mathcal{E}^o$ is defined as:
\begin{equation}\label{e:geo_graph_edge}
    e^o_{i, j} = \mathrm{exp} \left({- dist(i,j)}\right) ,
\end{equation}
where $dist(\cdot,\cdot)$ denotes the distance between POI i and j. 
The adjacent matrix $\mathbf{A}_{geo}$ can be constructed based on the edge constraints between each pair of out-of-town POIs.

Recently, GNN has been proved to be effective in modeling spatial data~\cite{zhang2020semi,li2020competitive,geng2019spatiotemporal}.
To capture the relations among POIs in a spatial perspective, we employ the graph neural network~\cite{kipf2016semi} as below:
\begin{equation}\label{e:geo_conv}
    {\mathbf{V}^o}^{\prime} = 
    \mathrm{ReLU}\left(
    \mathbf{A}_{geo} \mathbf{V}^o \mathbf{W}^c + \bm{b}^c
    \right)  ,
\end{equation}
where $\mathbf{W}^c \in \mathbb{R}^{d \times d}$ is a transition matrix and $\bm{b}^c \in \mathbb{R}^d$ is a bias vector.
${\mathbf{V}^o}^{\prime} = 
{\left(
{\mathbf{v}^o_1}^{\prime},{\mathbf{v}^o_2}^{\prime}, \ldots,
{\mathbf{v}^o_{D_2}}^{\prime}
\right)}^\mathrm{T}$
is the updated out-of-town POI embedding matrix,
which encodes geographical influence of POIs.

Moreover, from the users' point of view,
we adopt the matrix factorization (MF) method to explore the interactions between users and POIs in out-of-town.
In particular, we first assign a $d$-dimensional embedding, denoted by $\mathbf{u}^o \in \mathbb{R}^d$, to each of the users who left out-of-town check-ins.

Then, we aggregate the user's out-of-town preference and travel intention:
\begin{equation}\label{e:fuse}
    {\mathbf{u}^o}^{\prime} = 
    \mathrm{ReLU}\left(
    \mathbf{W}^f \mathrm{concat} \left(
        \mathbf{u}^o, \mathbf{u}^{(int)}
    \right)
    + \bm{b}^f
    \right) ,
\end{equation}
where $\mathbf{W}^f \in \mathbb{R}^{d \times 2d}$ is a transition matrix, $\bm{b} \in \mathbb{R} ^ d$ is a bias vector, and
$\mathrm{concat}(\cdot, \cdot)$ is a function concatenating its two input vectors.

Afterward, following the idea of MF that a user's scores over POIs can be regarded as the inner product of the user's latent embedding and the POIs', we define the score of user $i$ over out-of-town POI $j$ as follows:
\begin{equation}\label{e:score}
    s(i,j) = 
    \left({\mathbf{u}^o_i}^{\prime}\right)^\mathrm{T}
    {\mathbf{v}^o_j}^{\prime} .
\end{equation}

At last, following the assumption of BPR~\cite{rendle2012bpr} that the observed items should be ranked higher than those unobserved, for each user $u$, we randomly select a fixed size of positive samples visited by $u$ and their counterparts not checked in by $u$. 
Based on pairwise comparisons, the out-of-town preference loss is given by:
\begin{equation}\label{e:bpr_loss}
    \mathcal{L}_P = - 
    \sum_{u \in \mathcal{U}}
    \sum_{j \in \vec{c}_o}
    \sum_{k \notin \vec{c}_o} 
    \log\zeta\left( s(i,j) - s(i,k) \right) ,
\end{equation}
where $\vec{c}_o$ comprises $u$'s out-of-town check-ins.

\subsection{Preference Transfer}
Inspired by \cite{man2017cross}, we adopt an MLP as the non-linear mapping function to transfer user's home-town preference to out-of-town check-in bahavior.
We define the preference transfer loss as follows:
\begin{equation}\label{e:transfer_loss}
    \mathcal{L}_T = 
    \sum_{i \in \mathcal{U}}
     ||F_{tr}\left(\mathbf{u}^h_i\right) - \mathbf{u}^o_i||^2 ,
\end{equation}
where $F_{tr}$ is the MLP-based mapping function.

\subsection{Joint Training and Recommendation}
By combining the intention inference loss in~\cref{e:ntm_loss}, the preference loss in~\cref{e:bpr_loss} and the transfer loss in \cref{e:transfer_loss}, we can minimize the following composite loss function to jointly train our model in an end-to-end fashion:
\begin{equation}\label{e:joint_loss}
    \mathcal{L} = \lambda_1 \mathcal{L}_N + 
    \lambda_2 \mathcal{L}_P +
    \lambda_3 \mathcal{L}_T ,
\end{equation}
where $\lambda_1$, $\lambda_2$ and $\lambda_3$ are three hyper-parameters that control the respective contributions to the composite loss function.

After the parameters in our model are optimized, we can make recommendations for out-of-town users.
Specifically, given a user $u^{\ast} \notin \mathcal{U}$ and his/er home-town check-ins, we first generate his/er affine out-of-town user preference by using the trained preference transfer:
\begin{equation}\label{e:affine}
    \widehat{\mathbf{u}^o_{\ast}} = F_{tr}(\mathbf{u}^h_{\ast}) ,
\end{equation}
where $\mathbf{u}^h_{\ast}$ is $u^{\ast}$'s home-town preference embedding obtained from~\cref{e:home_preference}.
Meanwhile, we can obtain his/er intention embedding $\mathbf{u}^{(int)}_{\ast}$ by using~\cref{e:intention_embedding}. 
Similar to~\cref{e:fuse}, the travel intention embedding can be calculated as:
\begin{equation}\label{e:fuse_test}
    \widehat{{\mathbf{u}^o_{\ast}}^{\prime}} = 
    \mathrm{ReLU}\left(
    \mathbf{W}^f \mathrm{concat} \left(
        \widehat{\mathbf{u}^o_{\ast}}, \mathbf{u}^{(int)}_{\ast}
    \right)
    + \bm{b}^f
    \right) .
\end{equation}

Then, with $\widehat{{\mathbf{u}^o_{\ast}}^{\prime}}$ and ${\mathbf{V}^o}^{\prime}$, we can estimate the score of user $u^{\ast}$ over out-of-town POI $j$:
\begin{equation} \label{e:rec_score}
    \widehat{s({\ast},j)} = 
    \left(\widehat{{\mathbf{u}^o_{\ast}}^{\prime}}\right)^\mathrm{T}
    {\mathbf{v}^o_j}^{\prime} .
\end{equation}

Finally, we can pick the top-k out-of-town POIs based on the estimated scores as the recommendations for the out-of-town user $u^{\ast}$.

\section{Experiments}

\subsection{Experimental Setups}

\subsubsection{Dataset.} 
We chose three real-world travel behavior datasets including \bjsh, \shhz ~and \gzfs, to evaluate our approach.
\bjsh ~stands for traveling from Beijing to Shanghai, \shhz ~for Shanghai to Hangzhou and \gzfs ~for Guangzhou to Foshan. 
The travel records of the above three datasets were generated between 07/01/2019 and 12/31/2019. 
To ensure the data quality, in each dataset, we filtered out the POIs that is visited less than 5 times. 
Besides, the users, whose home-town check-ins are less than 5 or out-of-town check-ins are less than 3, were eliminated. 
Then, we randomly split users following the proportions: 80\%, 10\%, and 10\% to form a training set, a test set, and a validation set. 
%
The statistics of our dataset are given in~\cref{tab:dataset}.
Notably, in our datasets, each user has only one travel record, which guarantees the fairness of our evaluations for out-of-town recommendation. 

\begin{table}[t] 
    \tabcolsep 0.06in
    \caption{Basic description of datasets.} \label{tab:dataset}
    \begin{center}
    \begin{tabular}{ c|c|c|c|c }
      \toprule
      \multicolumn{2}{c|}{Dataset}   & \# Users  & \# POIs   & \# Check-ins  
      \\ \hline \hline 
      \multirow{2}{*}{\bjsh} 
        & Beijing   & \multirow{2}{*}{10,776}   & 2,111 & 127,528   
      \\ \cline{2-2}\cline{4-5}
        & Shanghai  &                           & 1,140 & 70,794    
      \\ \hline
        \multirow{2}{*}{\shhz} 
        & Shanghai  & \multirow{2}{*}{19,997}   & 3,415 & 263,158   
      \\ \cline{2-2}\cline{4-5}
        & Hangzhou  &                           & 1,203 & 116,475   
      \\ \hline
      \multirow{2}{*}{\gzfs} 
        & Guangzhou & \multirow{2}{*}{12,788}   & 4,228 & 220,006   
      \\ \cline{2-2}\cline{4-5}
        & Foshan    &                           & 1,225 & 57,229    
      \\ 
      \bottomrule
  \end{tabular}
  \end{center}
\end{table}

\subsubsection{Evaluation Metrics.}
Since there are more than one out-of-town check-ins (i.e. multiple ground-truths) for each user in our dataset, 
we apply \textit{Recall@k}~(Rec@k) and \textit{mean average precision}~(MAP) to evaluate the performance of different recommender systems. 
The larger the values of the above metrics are, the better the models perform.

\subsubsection{Baselines.}
We compared our approach with various baselines that could be used for out-of-town recommendation.

\begin{itemize}
  \item 
  \textbf{TOP} is a naive method which recommends the top-N frequently visited POIs in the target city.
  \item 
  \textbf{UCF} is a user-based collaborative filtering method which recommends POIs for a target user in accordance with POI check-in behaviors of similar users. 
  \item 
  \textbf{BPR-MF}~\cite{rendle2012bpr} takes MF as the underlying predictor, which aims to factorize the user-POI matrix into the latent factors, and optimizes the MF by Bayesian Personalized Ranking~(BPR). 
  Recommendations are implemented based on the reconstruction of the matrix. 
  \item 
  \textbf{GRU4Rec}~\cite{hidasi2015session} utilizes RNNs to model users' sequential check-ins. 
  To make this method capable of our problem, we take the home-town check-ins as RNNs' input, predict the out-of-town check-ins by utilizing the hidden state, and train the model by BPR.
  \item 
  \textbf{SR-GNN}~\cite{wu2019session} utilizes GNNs to model the complex transitions of items. 
  Similar to GRU4Rec, we regard each user's home-town check-ins as a directed graph, predict the out-of-town check-ins and train the model using BPR.
  \item 
  \textbf{LA-LDA}~\cite{yin2014lcars} is a location-aware recommendation model which is suitable for out-of-town recommendation scenario. 
  It takes personal interests and geographical gaps into consideration by exploiting POI co-visiting patterns.
  \item 
  \textbf{EMCDR}~\cite{man2017cross} is a cross-domain recommendation approach, which uses a multi-layer perceptron to capture the nonlinear mapping function across domains.
\end{itemize}

Moreover, to explore the respective contributions of different modules in our approach, we further come up with three variants of \ourmodel~as follows:

\begin{itemize}
    \item 
    \textbf{\ourmodel-I}: this variant removes travel intention discovery module. 
    As a result, it recommends only based on users' preference.
    \item 
    \textbf{\ourmodel-C}: this variant removes the GeoConv, such that the geographical influence of out-of-town POIs is neglected.
    \item 
    \textbf{\ourmodel-IC}: this variant removes both travel intention discovery module and the geographical influence.
\end{itemize}

\subsubsection{Implementations.}
The number $d$ (i.e. the hidden size) was fixed to 128 for all latent representations. 
In the travel intention discovery module, we set the topic number $K$ as 15 for better explanation. 
In the joint training stage, we set $\lambda_1=\lambda_2=\lambda_3=1$ in \cref{e:joint_loss}. 
We used Adam optimizer to train our approach with an initial learning rate as 0.001 and an L2 regularization with weight $10^{-5}$.
When the quantity measures were evaluated, the test was repeated over 5 times using different data splits and the average was reported.

\subsection{Experimental Results}

\subsubsection{Recommendation Performance.}

\begin{table*}[t] 
\renewcommand\arraystretch{1}
\tabcolsep 0.025in
\caption{The overall performance of \ourmodel~and baselines.}
\label{tab:performance}
\begin{center}
  \begin{tabular}{c||c|c|c|c|c|c|c|c|c|c|c|c}
      \hline
      \multirow{2}*{Methods} & \multicolumn{4}{c}{\bjsh} & 
      \multicolumn{4}{|c}{\shhz} &
      \multicolumn{4}{|c}{\gzfs} \\
      \cline{2-13}
      ~ & Rec@10 & Rec@20 & Rec@30 & MAP & Rec@10 & Rec@20 & Rec@30 & MAP & Rec@10 & Rec@20 & Rec@30 & MAP \\
      \hline
      \hline
      LA-LDA &
      0.0160 & 0.0335 & 0.0417 & 0.0151 &
      0.0008 & 0.0021 & 0.0028 & 0.0019 &
      0.0020 & 0.0036 & 0.0057 & 0.0021 \\
      UCF & 
      0.0443 & 0.0700 & 0.0935 & 0.1133 &
      0.0628 & 0.0874 & 0.0981 & 0.2577 &
      0.0386 & 0.0661 & 0.0800 & 0.1071 \\
      SR-GNN &
      0.1168 & 0.1807 & 0.2627 & 0.1071 &
      0.2287 & 0.4550 & 0.5661 & 0.2013 &
      0.0933 & 0.1670 & 0.2541 & 0.0566 \\
      BPR-MF & 
      0.1768 & 0.2379 & 0.2844 & 0.0901 &
      0.2812 & 0.3588 & 0.4116 & 0.1910 &
      0.1642 & 0.2545 & 0.3173 & 0.0947 \\
      TOP &
      0.2062 & 0.3103 & 0.3818 & 0.1494 &
      0.3713 & 0.4620 & 0.5176 & 0.2896 &
      0.1964 & 0.2838 & 0.3483 & 0.1202 \\
      GRU4Rec &
      0.2091 & 0.3011 & 0.3763 & 0.1438 &
      0.3619 & 0.4650 & 0.5150 & 0.2807 & 
      0.1789 & 0.2742 & 0.3422 & 0.1034 \\
      EMCDR & 
      0.2163 & 0.3008 & 0.3649 & \textbf{0.1553} &
      0.3772 & 0.4358 & 0.4732 & \textbf{0.3260} &
      0.1928 & 0.2770 & 0.3368 & 0.1246 \\
      \hline
      \ourmodel-IC &
      0.2029 & 0.2880 & 0.3513 & 0.1497 &
      0.3679 & 0.4406 & 0.4963 & 0.3020 &
      0.1937 & 0.2609 & 0.3178 & 0.1245 \\
      \ourmodel-I &
      0.2177 & 0.3084 & 0.3825 & 0.1543 &
      0.3825 & 0.4624 & 0.5177 & 0.3016 &
      0.2028 & 0.2841 & 0.3449 & \textbf{0.1266} \\
      \ourmodel-C &
      \textbf{0.2233} & 0.3194 & \textbf{0.3955} & 0.1538 &
      \textbf{0.3914} & 0.4757 & \textbf{0.5300} & 0.2950 &
      0.2032 & 0.2918 & \textbf{0.3569} & 0.1246 \\
      \ourmodel &
      0.2226 & \textbf{0.3198} & 0.3938 & 0.1541 &
      \textbf{0.3914} & \textbf{0.4768} & 0.5295 & 0.2955 &
      \textbf{0.2039} & \textbf{0.2922} & 0.3551 & 0.1246 \\
      \hline
  \end{tabular}
\end{center}
\end{table*}

The performances of \ourmodel ~as well as its variants and the baselines are illustrated in \cref{tab:performance}. 
Basically, \ourmodel~consistently outperforms the baselines w.r.t. Rec@k.
Regarding MAP, \ourmodel~achieves best performance on \gzfs ~dataset, and second best on \bjsh ~and \shhz ~datasets.

In particular, UCF, LA-LDA and BPR-MF perform relatively worse. 
UCF and BPR-MF are two collaborative filtering algorithms for item recommendation, 
which cannot be directly applied to out-of-town recommendation due to the data scarcity issues. 
Though LA-LDA considers the geographical gaps, it is still insufficient to model the fine-grained personal interest drifts when the difference is big~(e.g. cross city), which makes it less competitive for out-of-town recommendation. 

TOP is not a personalized method and makes recommendations only based on the popularity of POIs according to history logs, 
yet has surprisingly better performances than some personalized approaches. 
The probable reason is that the out-of-town travel behaviors are usually dominated by tourism, which makes some hot attractions~(e.g. famous landmarks) be frequently visited by the out-of-town users.

GRU4Rec and SR-GNN are two session-based deep recommender systems that take sequential and structural information into account, respectively. 
However, they also neglect the users' interest drifts and context differences between home-town areas and out-of-town areas.

EMCDR is the state-of-the-art cross-domain recommendation framework. 
Because of its capability of non-linear mapping function that transfers features from the source domain to the target domain, EMCDR achieves the almost best ranking performance when making out-of-town recommendations.
One possible reason is that in our \ourmodel~framework, negative sampling strategy is not adopted in home-town preference modeling compared to EMCDR, 
which may lead to higher ranking of some negative items in home-town area and may have a negative impact on ranking performance, i.e. MAP. 
%
However, \ourmodel~outperforms EMCDR with satisfactory margins in terms of Rec@k, which indicates that \ourmodel~is more effective in retrieving relevant out-of-town POIs and more beneficial for out-of-town recommendation in practice.

\subsubsection{Ablation Analysis.}
As for the variants of \ourmodel, we have the following main observations:

1) \ourmodel~outperforms \ourmodel-I w.r.t. Rec@k, which 
indicates the effectiveness of taking into account users' out-of-town intentions. 
Besides, the MAP slightly falls when travel intention discovery module is utilized. 
The reason might be that global signals such as intentions can become disturbance while the model is trying to put every item in a right ranking.

2) With comparing the results between \ourmodel-IC and \ourmodel-I, 
the removal of GeoConv decreases the performances on all metrics. 
However, we also find that with the existence of travel intention discovery module, GeoConv barely contributes to the performances, since GeoConv may result in overfitting as the learned intention embedding contains potential relations between POIs.

\subsubsection{Case Study on Intention Discovery.}

\begin{figure}[t]
\centering
\subfigure[The distributions of visited POIs (over generic intentions).]{
\includegraphics[width=0.95\columnwidth]{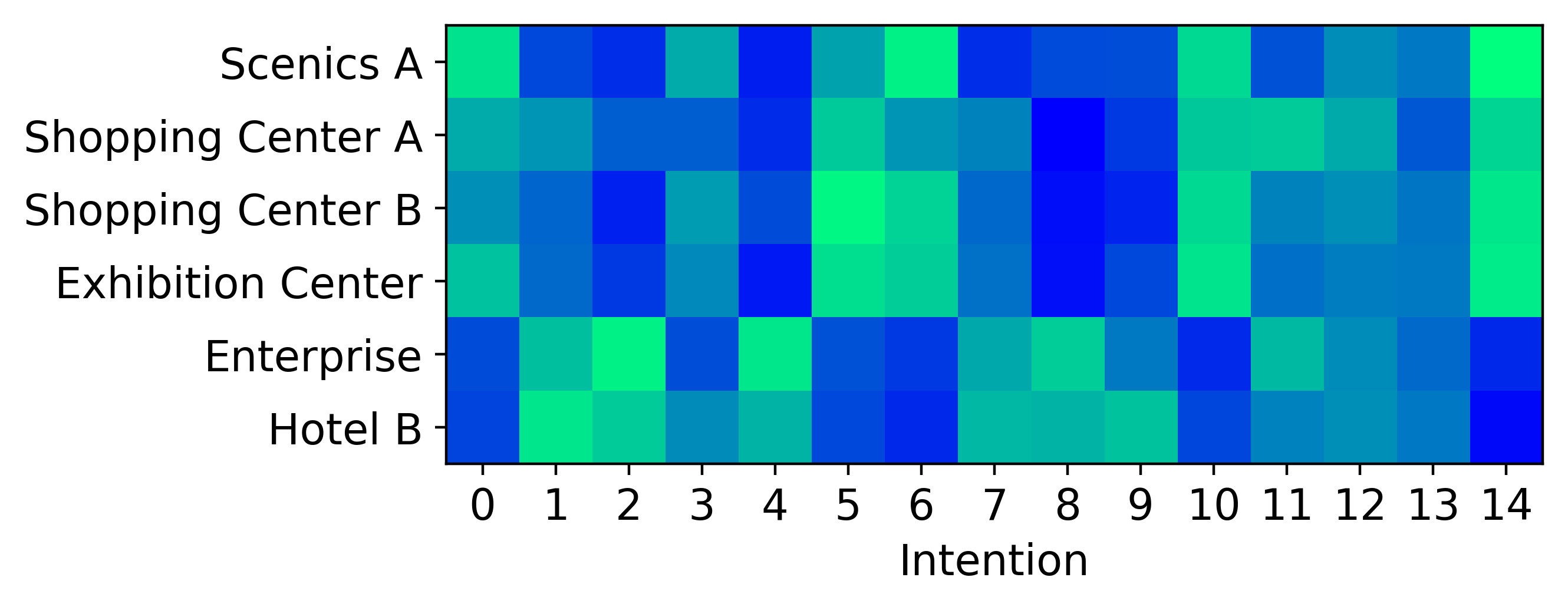}
}
\subfigure[The weights of generic intentions for user-specific intentions.]{
\includegraphics[width=0.95\columnwidth]{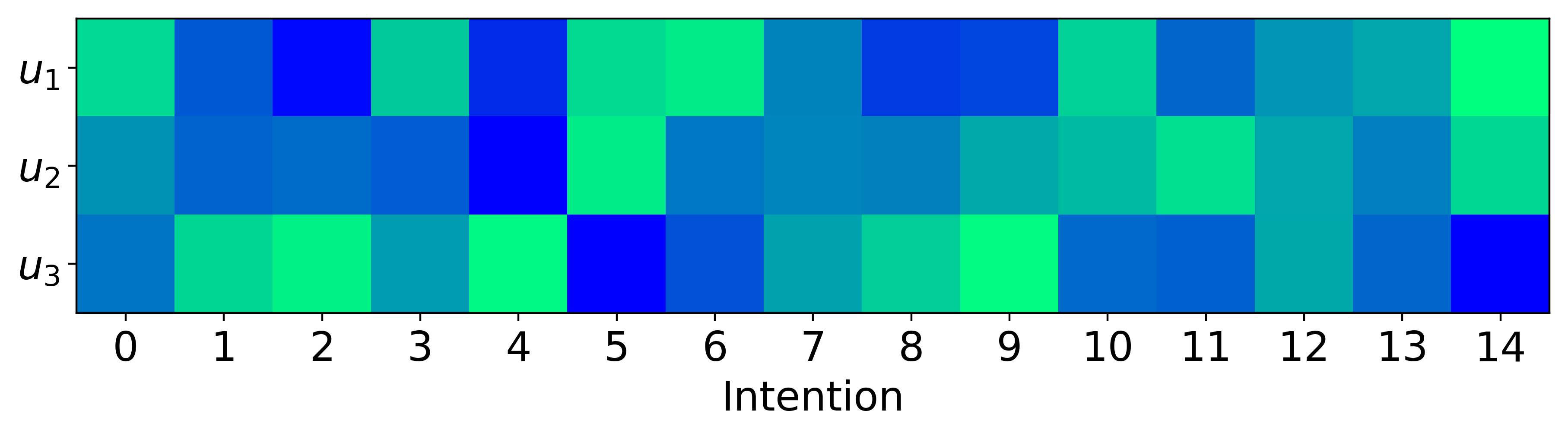}
}
\caption{The visualization of the case study.}
\label{fig:word}
\end{figure}

\newcommand{\tabincell}[2]{\begin{tabular}{@{}#1@{}}#2\end{tabular}}
\begin{table}
\renewcommand\arraystretch{1}
\tabcolsep 0.05in
\caption{Out-of-town check-ins of three selected users from test set.}
\label{tab:user_case}
\begin{center}
  \begin{tabular}{c|c}
      \toprule
      User & Out-of-town check-ins \\
      \hline
      \hline
      {$u_1$} & 
      \tabincell{c}{
      Scenics A, Scenics B, \\
      Art Gallery, Shopping Center A
      } \\
      \hline
      {$u_2$} & 
      \tabincell{c}{
      Shopping Center B, Exhibition Center, \\
      Life Plaza, Shopping Center C, Hotel A} \\
      \hline
      {$u_3$} & 
      \tabincell{c}{
      Enterprise, Hotel B, Hotel C
      } \\
      \bottomrule
  \end{tabular}
\end{center}
\end{table}

We next present a case study on the discovered intentions to further evaluate \ourmodel~framework.
We randomly selected 3 recommended cases with promising Rec@30 (e.g. 0.67 for $u_1$, 0.33 for $u_2$ and 0.5 for $u_3$) from \bjsh~dataset.
The out-of-town check-ins of these cases are illustrated in \cref{tab:user_case}.
Besides, we visualized the POI-intention distributions~(i.e. $\Phi^\mathrm{T}$) of some POIs visited by $u_1$, $u_2$ and $u_3$ in \cref{fig:word}(a),
and the weights of generic intentions for the user-specific attentions (refer to \cref{e:intention_embedding}) related to these users in \cref{fig:word}(b). 
The deeper the color, the greater the value.

As depicted in \cref{tab:user_case}, we can infer that $u_1$ traveled to Shanghai for vacation, $u_2$ for shopping and $u_3$ for business, respectively.
Besides, based on the inference of users' intentions, we can clearly tell that the difference between $u_1$ and $u_2$ is small, while, the difference between $u_3$ and $u_1$/$u_2$ is large, regarding the travel intention (refer to \cref{fig:word}(b)). 

Moreover, the intention distributions of the representative POIs are also distinguishable (refer to \cref{fig:word}(a)).
For example, the distributions of scenics and shopping centers have more weights on intention \#2, \#4, \#8, etc., whereas the distributions of functional facilities~(e.g. hotels and enterprises) have more weights on intention \#5, \#6, \#10, \#14, etc. 
Hence, we can conclude that the former set of intentions is more leisure-related while the latter is more function-related.

\subsubsection{Parameter Sensitivity.}

\begin{figure}[t]\label{fig:sensitivity}
  \centering
  \subfigure[Effect of $K$]{
    \includegraphics[width=0.45\columnwidth]{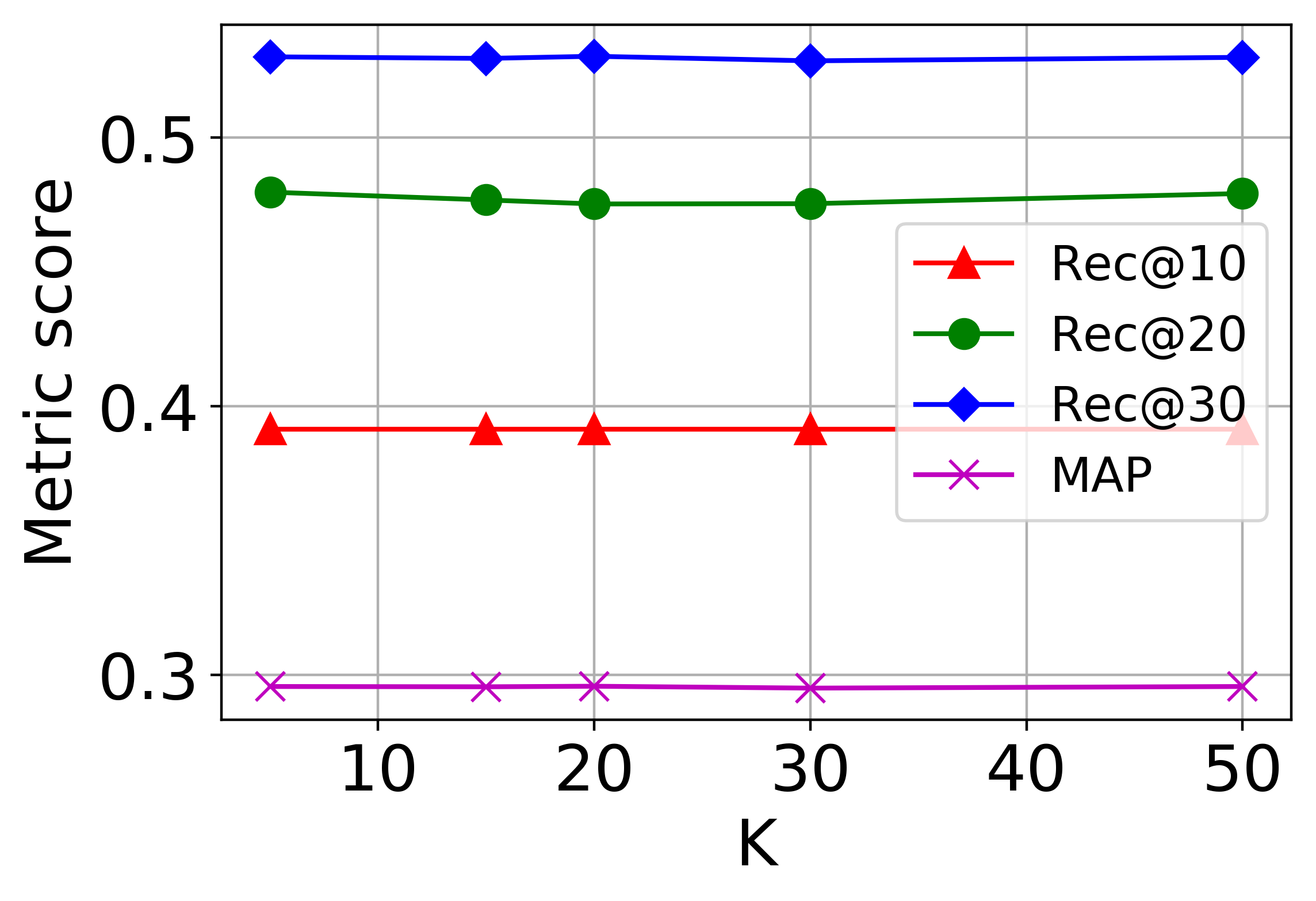}
  }
  \subfigure[Effect of $\lambda_1$]{
    \includegraphics[width=0.45\columnwidth]{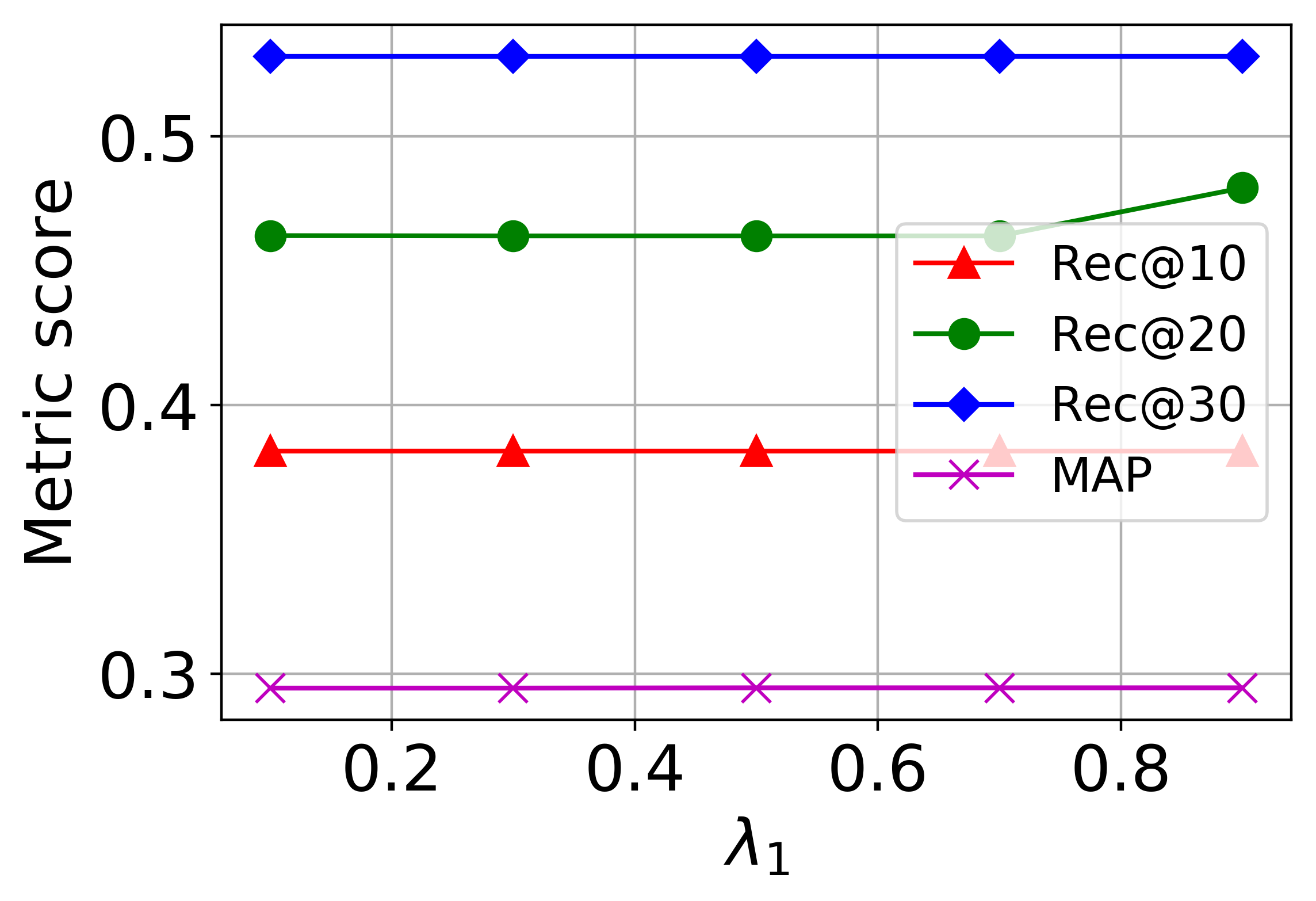}
  }
  \subfigure[Effect of $\lambda_2$]{
    \includegraphics[width=0.45\columnwidth]{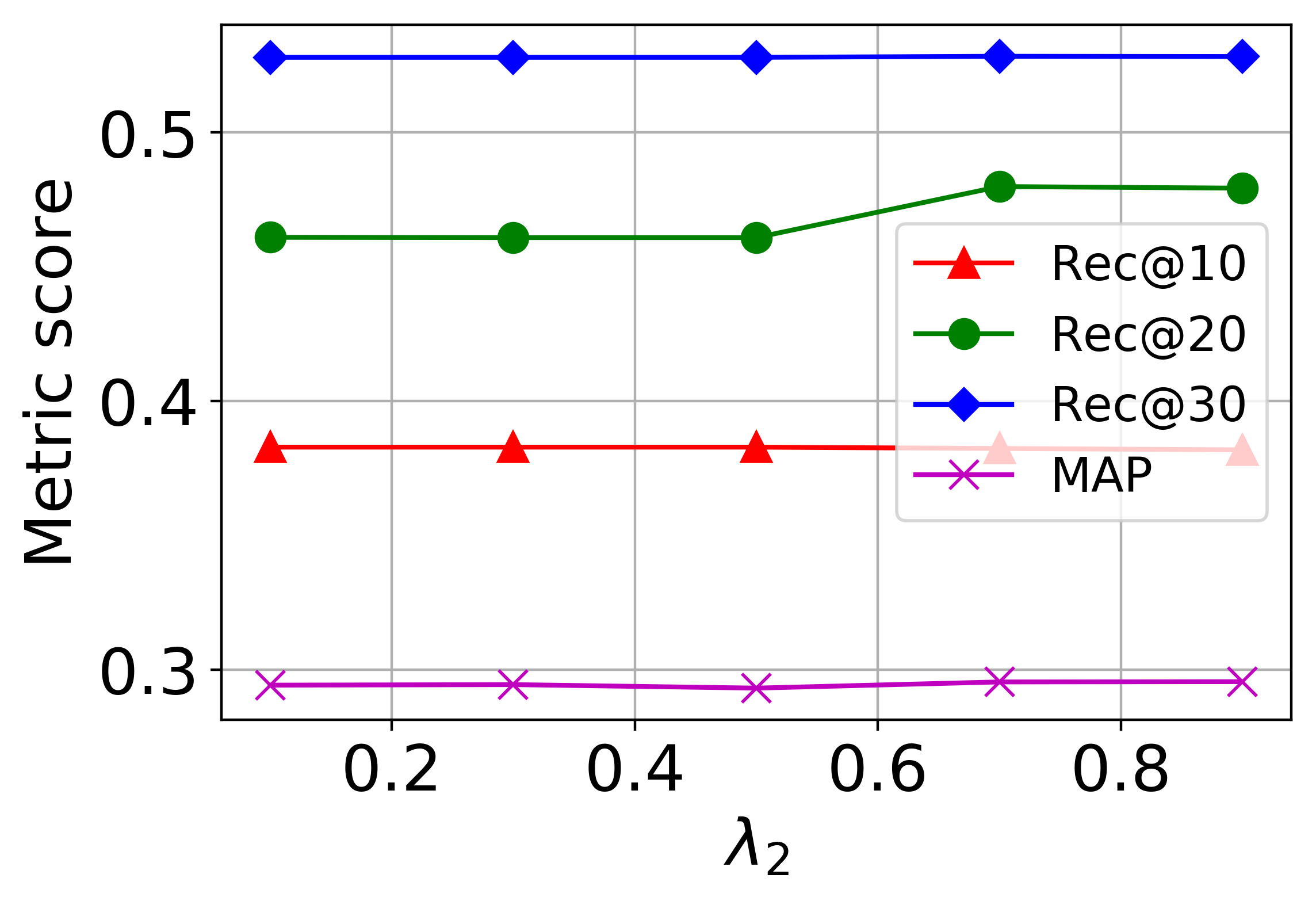}
  }
  \subfigure[Effect of $\lambda_3$]{
    \includegraphics[width=0.45\columnwidth]{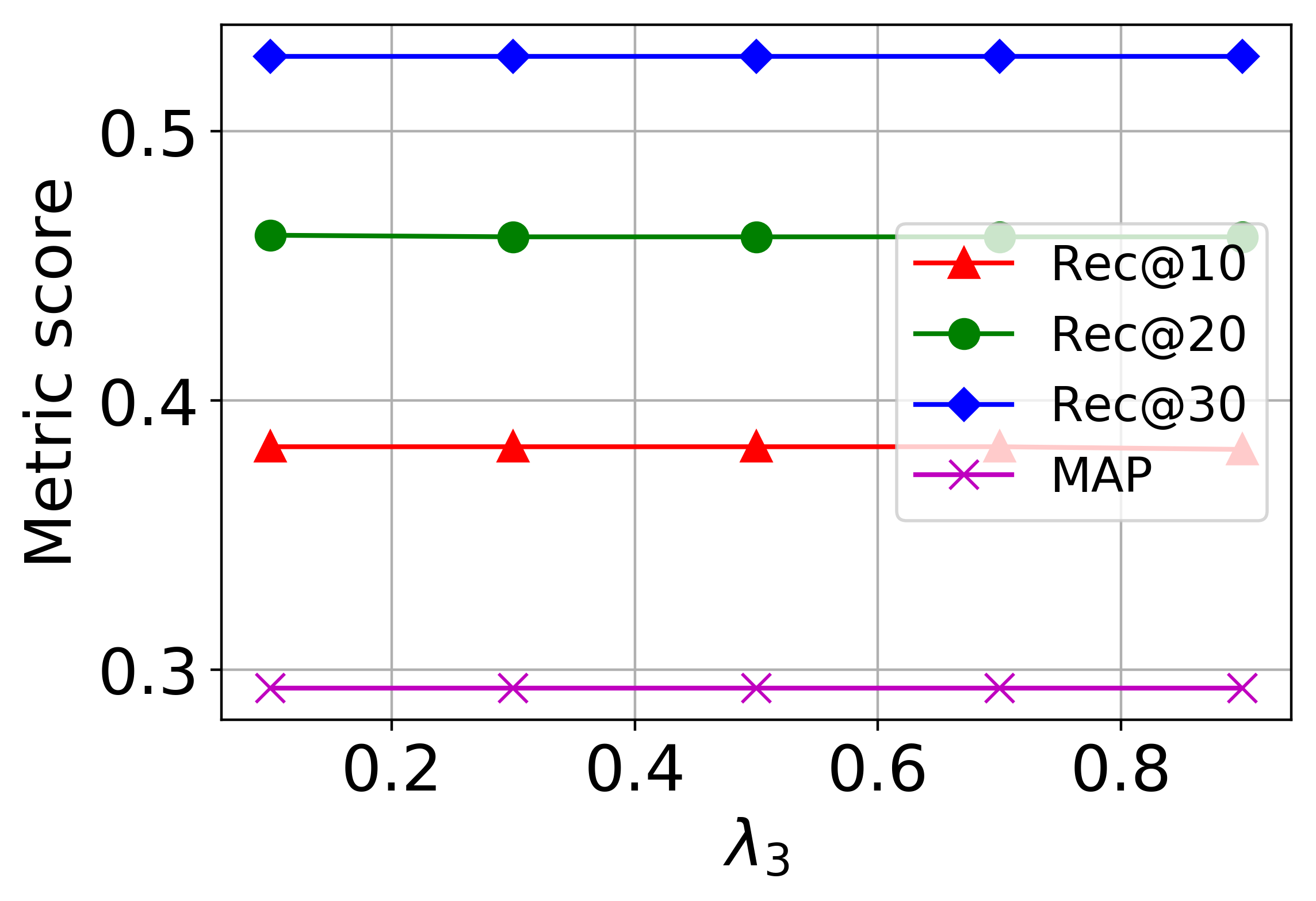}
  }
  \caption{The impacts of inherent intention number ($K$) and different loss function weights ($\lambda_1$, $\lambda_2$ and $\lambda_3$) on \shhz~dataset w.r.t. the recommendation performance.}
  \label{fig:parameter-sensitive}
\end{figure}

We report the influence of the number of inherent travel intentions, i.e. $K$, and the impacts of $\lambda_1$, $\lambda_2$ and $\lambda_3$ in loss functions, respectively. 
We only demonstrate the results on \shhz~dataset. 
The results are similar on the other two datasets. 
Besides, to evaluate the impacts of parameters $\lambda_1$, $\lambda_2$ and $\lambda_3$ on the recommendation performance, we first vary $\lambda_1$ while setting $\lambda_2 = \lambda_3 = \frac{1-\lambda_1}{2}$, and we use the similar strategy for evaluating $\lambda_2$ and $\lambda_3$, respectively. 

As shown in \cref{fig:parameter-sensitive}(a), the scores are relatively stable as $K$ increases. 
The reason behind is that the user-specific intention is a summary of all generic intention embeddings, which makes the overall performance insensitive to the number of generic inherent intentions. 
Moreover, from Figs.~\ref{fig:parameter-sensitive}(b) to \ref{fig:parameter-sensitive}(d), we can observe that, 
with increasing $\lambda_1$, $\lambda_2$ and $\lambda_3$ from 0.1 to 0.9, both the Recall and MAP metrics are stable in general with an exception that the Rec@20 is slightly better with larger $\lambda_1$ and $\lambda_2$.

\section{Related Work}

{\bf{Out-of-town POI recommendation}}
attempts to provide a list of POIs out-of-town users are interested in~\cite{wang2017location}. 
To tackle the out-of-town recommendation problem, many informative features had been investigated, including user preference~\cite{ference2013location}, geographical influence of out-of-town check-ins~\cite{ference2013location,pham2017general}, and social influence~\cite{ference2013location}.
One important phenomenon that out-of-town recommendations should not neglect is user's interest drift, i.e. user's out-of-town check-ins are not aligned to user's home-town check-in preference. 
Some studies have paid attentions to make out-of-town recommendation by taking into account user's interest drift~\cite{yin2014lcars,yin2016joint,wang2017location}. 
Most of them take the textual reviews as input with employing the topic model. 
However, the data sparsity issue is getting worse when utilizing the textual content related to POIs and users.
Besides, as a number of POIs in out-of-town check-ins relate to tourism, there also exist researches focusing on out-of-town tourism POI recommendation~\cite{liu2011personalized,brilhante2013shall,hu2017unifying}. 
Our work differentiates itself from previous works by comprehensively considering user's preference, travel intention, geographical constraints and user interest drifts for out-of-town recommendation.

{\bf{Topic models}}
%
have been widely applied as generative models for different tasks \cite{wang2017location,xu2017measuring,shen2018joint,zhou2019topic,luo2020spatial}.
However, as the dimensionality grows, these methods are scant to perform fast and accurate inference. 
Recently, deep learning techniques and neural variational inference have accelerated the development of latent variable models~\cite{miao2016neural,miao2017discovering,kingma2013auto,srivastava2017autoencoding}. 
For example, \cite{miao2016neural} proposes a neural variational document model (NVDM) for text mining. 
Besides, Neural Topic Model (NTM)~\cite{miao2017discovering} was proposed to discover latent topics by variational inference. 
These methods provide us new opportunities to tackle topic discovery problem in a data driven fashion.

\section{Concluding Remarks}

In this paper, we studied the out-of-town recommendation problem via travel intention modeling. 
We proposed a data-driven framework \ourmodel~to learn an out-of-town recommender by comprehensively taking user preference, interest drifts, travel intention and out-of-town geographical influence into account as a whole. 
To investigate user's home-town preference, a G-GNN model was exploited.
On the other hand, the user's out-of-town preference was estimated in a collective manner and enriched through a geographical GCN. 
Afterward, we devised a preference transfer module to map home-town preference to out-of-town check-in behavior via an MLP. 
Moreover, to understand the user's complex travel intention, we developed an NTM based user-specific travel intention discovery module. 
Finally, with jointly minimizing composite loss, \ourmodel~can yield the learned recommender in an end-to-end fashion.
Through extensive experiments on real-world datasets, we validated the effectiveness of \ourmodel~quantitatively. 
A case study further validated the ability of \ourmodel~to understand users' travel intentions.

\section*{Acknowledgments}

This work is supported in part by NSFC {71531001, 61725205, 91746301, and 61703386}.

\bibliography{refs}

\begin{thebibliography}{28}
\providecommand{\natexlab}[1]{#1}
\providecommand{\url}[1]{\texttt{#1}}
\providecommand{\urlprefix}{URL }
\expandafter\ifx\csname urlstyle\endcsname\relax
  \providecommand{\doi}[1]{doi:\discretionary{}{}{}#1}\else
  \providecommand{\doi}{doi:\discretionary{}{}{}\begingroup
  \urlstyle{rm}\Url}\fi

\bibitem[{Brilhante et~al.(2013)Brilhante, Macedo, Nardini, Perego, and
  Renso}]{brilhante2013shall}
Brilhante, I.; Macedo, J.~A.; Nardini, F.~M.; Perego, R.; and Renso, C. 2013.
\newblock Where shall we go today? Planning touristic tours with TripBuilder.
\newblock In \emph{CIKM'13}, 757--762.

\bibitem[{Cho et~al.(2014)Cho, Van~Merri{\"e}nboer, Bahdanau, and
  Bengio}]{cho2014properties}
Cho, K.; Van~Merri{\"e}nboer, B.; Bahdanau, D.; and Bengio, Y. 2014.
\newblock On the properties of neural machine translation: Encoder-decoder
  approaches.
\newblock \emph{arXiv preprint arXiv:1409.1259} .

\bibitem[{Ference, Ye, and Lee(2013)}]{ference2013location}
Ference, G.; Ye, M.; and Lee, W.-C. 2013.
\newblock Location recommendation for out-of-town users in location-based
  social networks.
\newblock In \emph{CIKM'13}, 721--726.

\bibitem[{Geng et~al.(2019)Geng, Li, Wang, Zhang, Yang, Ye, and
  Liu}]{geng2019spatiotemporal}
Geng, X.; Li, Y.; Wang, L.; Zhang, L.; Yang, Q.; Ye, J.; and Liu, Y. 2019.
\newblock Spatiotemporal multi-graph convolution network for ride-hailing
  demand forecasting.
\newblock In \emph{AAAI'19}, volume~33, 3656--3663.

\bibitem[{Hidasi et~al.(2015)Hidasi, Karatzoglou, Baltrunas, and
  Tikk}]{hidasi2015session}
Hidasi, B.; Karatzoglou, A.; Baltrunas, L.; and Tikk, D. 2015.
\newblock Session-based recommendations with recurrent neural networks.
\newblock \emph{arXiv preprint arXiv:1511.06939} .

\bibitem[{Hu et~al.(2017)Hu, Shao, Shen, Huang, and Shen}]{hu2017unifying}
Hu, G.; Shao, J.; Shen, F.; Huang, Z.; and Shen, H.~T. 2017.
\newblock Unifying multi-source social media data for personalized travel route
  planning.
\newblock In \emph{SIGIR'17}, 893--896.

\bibitem[{Kingma and Welling(2013)}]{kingma2013auto}
Kingma, D.~P.; and Welling, M. 2013.
\newblock Auto-encoding variational bayes.
\newblock \emph{arXiv preprint arXiv:1312.6114} .

\bibitem[{Kipf and Welling(2016)}]{kipf2016semi}
Kipf, T.~N.; and Welling, M. 2016.
\newblock Semi-supervised classification with graph convolutional networks.
\newblock \emph{arXiv preprint arXiv:1609.02907} .

\bibitem[{Li et~al.(2020)Li, Zhou, Xu, Liu, Lu, and Xiong}]{li2020competitive}
Li, S.; Zhou, J.; Xu, T.; Liu, H.; Lu, X.; and Xiong, H. 2020.
\newblock Competitive Analysis for Points of Interest.
\newblock In \emph{KDD'20}, 1265--1274.

\bibitem[{Li et~al.(2015)Li, Tarlow, Brockschmidt, and Zemel}]{li2015gated}
Li, Y.; Tarlow, D.; Brockschmidt, M.; and Zemel, R. 2015.
\newblock Gated graph sequence neural networks.
\newblock \emph{arXiv preprint arXiv:1511.05493} .

\bibitem[{Liu et~al.(2011)Liu, Ge, Li, Chen, and Xiong}]{liu2011personalized}
Liu, Q.; Ge, Y.; Li, Z.; Chen, E.; and Xiong, H. 2011.
\newblock Personalized travel package recommendation.
\newblock In \emph{ICDM'11}, 407--416. IEEE.

\bibitem[{Luo et~al.(2020)Luo, Zhou, Bao, Li, Culpepper, Ying, Liu, and
  Xiong}]{luo2020spatial}
Luo, H.; Zhou, J.; Bao, Z.; Li, S.; Culpepper, J.~S.; Ying, H.; Liu, H.; and
  Xiong, H. 2020.
\newblock Spatial object recommendation with hints: When spatial granularity
  matters.
\newblock In \emph{SIGIR'20}, 781--790.

\bibitem[{Man et~al.(2017)Man, Shen, Jin, and Cheng}]{man2017cross}
Man, T.; Shen, H.; Jin, X.; and Cheng, X. 2017.
\newblock Cross-Domain Recommendation: An Embedding and Mapping Approach.
\newblock In \emph{IJCAI'17}, 2464--2470.

\bibitem[{Miao, Grefenstette, and Blunsom(2017)}]{miao2017discovering}
Miao, Y.; Grefenstette, E.; and Blunsom, P. 2017.
\newblock Discovering discrete latent topics with neural variational inference.
\newblock \emph{arXiv preprint arXiv:1706.00359} .

\bibitem[{Miao, Yu, and Blunsom(2016)}]{miao2016neural}
Miao, Y.; Yu, L.; and Blunsom, P. 2016.
\newblock Neural variational inference for text processing.
\newblock In \emph{ICML'16}, 1727--1736.

\bibitem[{Pham, Li, and Cong(2017)}]{pham2017general}
Pham, T.-A.~N.; Li, X.; and Cong, G. 2017.
\newblock A general model for out-of-town region recommendation.
\newblock In \emph{WWW'17}, 401--410.

\bibitem[{Rendle et~al.(2012)Rendle, Freudenthaler, Gantner, and
  Schmidt-Thieme}]{rendle2012bpr}
Rendle, S.; Freudenthaler, C.; Gantner, Z.; and Schmidt-Thieme, L. 2012.
\newblock BPR: Bayesian personalized ranking from implicit feedback.
\newblock \emph{arXiv preprint arXiv:1205.2618} .

\bibitem[{Shen et~al.(2018)Shen, Zhu, Zhu, Xu, Ma, and Xiong}]{shen2018joint}
Shen, D.; Zhu, H.; Zhu, C.; Xu, T.; Ma, C.; and Xiong, H. 2018.
\newblock A joint learning approach to intelligent job interview assessment.
\newblock In \emph{IJCAI'18}, 3542--3548.

\bibitem[{Srivastava and Sutton(2017)}]{srivastava2017autoencoding}
Srivastava, A.; and Sutton, C. 2017.
\newblock Autoencoding variational inference for topic models.
\newblock \emph{arXiv preprint arXiv:1703.01488} .

\bibitem[{Wang et~al.(2017)Wang, Fu, Wang, Yin, Du, and
  Xiong}]{wang2017location}
Wang, H.; Fu, Y.; Wang, Q.; Yin, H.; Du, C.; and Xiong, H. 2017.
\newblock A location-sentiment-aware recommender system for both home-town and
  out-of-town users.
\newblock In \emph{SIGKDD'17}, 1135--1143.

\bibitem[{Wei and Mao(2019)}]{wei2019modeling}
Wei, P.; and Mao, W. 2019.
\newblock Modeling Transferable Topics for Cross-Target Stance Detection.
\newblock In \emph{SIGIR'19}, 1173--1176.

\bibitem[{Wu et~al.(2019)Wu, Tang, Zhu, Wang, Xie, and Tan}]{wu2019session}
Wu, S.; Tang, Y.; Zhu, Y.; Wang, L.; Xie, X.; and Tan, T. 2019.
\newblock Session-based recommendation with graph neural networks.
\newblock In \emph{AAAI'19}, volume~33, 346--353.

\bibitem[{Xu et~al.(2017)Xu, Zhu, Zhu, Li, and Xiong}]{xu2017measuring}
Xu, T.; Zhu, H.; Zhu, C.; Li, P.; and Xiong, H. 2017.
\newblock Measuring the popularity of job skills in recruitment market: A
  multi-criteria approach.
\newblock \emph{arXiv preprint arXiv:1712.03087} .

\bibitem[{Yin et~al.(2014)Yin, Cui, Sun, Hu, and Chen}]{yin2014lcars}
Yin, H.; Cui, B.; Sun, Y.; Hu, Z.; and Chen, L. 2014.
\newblock LCARS: A spatial item recommender system.
\newblock \emph{TOIS} 32(3): 1--37.

\bibitem[{Yin et~al.(2016)Yin, Cui, Zhou, Wang, Huang, and
  Sadiq}]{yin2016joint}
Yin, H.; Cui, B.; Zhou, X.; Wang, W.; Huang, Z.; and Sadiq, S. 2016.
\newblock Joint modeling of user check-in behaviors for real-time
  point-of-interest recommendation.
\newblock \emph{TOIS} 35(2): 1--44.

\bibitem[{Zeng et~al.(2018)Zeng, Li, Song, Gao, Lyu, and King}]{zeng2018topic}
Zeng, J.; Li, J.; Song, Y.; Gao, C.; Lyu, M.~R.; and King, I. 2018.
\newblock Topic memory networks for short text classification.
\newblock \emph{arXiv preprint arXiv:1809.03664} .

\bibitem[{Zhang et~al.(2020)Zhang, Liu, Liu, Zhou, and Xiong}]{zhang2020semi}
Zhang, W.; Liu, H.; Liu, Y.; Zhou, J.; and Xiong, H. 2020.
\newblock Semi-Supervised Hierarchical Recurrent Graph Neural Network for
  City-Wide Parking Availability Prediction.
\newblock In \emph{AAAI'20}, volume~34, 1186--1193.

\bibitem[{Zhou, Mascolo, and Zhao(2019)}]{zhou2019topic}
Zhou, X.; Mascolo, C.; and Zhao, Z. 2019.
\newblock Topic-enhanced memory networks for personalised point-of-interest
  recommendation.
\newblock In \emph{SIGKDD'17}, 3018--3028. ACM.

\end{thebibliography}

\end{document}